\documentclass[12pt,A4paper]{article}
\usepackage[T1]{fontenc}
\usepackage[utf8]{inputenc}
\usepackage[english]{babel}
\usepackage{fancyhdr}
\usepackage{amsmath}
\usepackage{amssymb}
\usepackage{amsfonts} 
\usepackage{graphicx}
\usepackage{anysize}
\usepackage{cancel}
\usepackage{psfrag}
\usepackage{secdot}
\usepackage{sectsty}
\usepackage{cite}
\usepackage{nameref}
\usepackage[verbose]{placeins}
\sectionfont{\fontsize{14}{15}\selectfont}
\subsectionfont{\fontsize{12}{15}\selectfont}

\begin{document}
		\marginsize{1.5cm}{0.5cm}{1cm}{1cm}
		
\begin{center}
	\mbox{\large\textbf{On the edge of complexity:}}\\
	\vspace{0.1cm}
	\mbox{\Large\textbf{The simplest not simple coupled mechanical system}}\\
	\vspace{0.2cm}	
	Gergely Nyitray\\
	\textit{nyitray.gergely@mik.pte.hu}\\
	\mbox{\small Department of Automation, Faculty of Engineering and Information Technology},\\ 
	\mbox{\small University of Pécs, Boszorkány street 2, Pécs, Hungary.}
\end{center}

\section*{Abstract}
\begin{small}		
We show that during normal modes of an oscillatory system consisting of a hoop and a cylinder joining their centers by an ideal spring, its central mass does not remain at rest. This effect is due to the resultant external static friction forces acting on the system which disappears when the coupled rigid bodies have the same moment of inertia. However, in case of different moment of inertia by proper positioning the two fixed end points of the spring vertically, it is shown that the central mass of the system remains at rest. The equation of motion of the coupled system is derived using dynamic equations, Lagrange's equations, Hamilton's equations and even by applying the conservation laws of energy and angular momentum. 
The relationship between the static friction forces acting on the rigid bodies is also examined.
\end{small}
\vspace{0.3cm}\\
\mbox{\small\emph{Keywords}: Oscillation; Coupling; Rolling of rigid bodies}

\section{Introduction}

\begin{figure}[h!]
	\begin{center}
		\begin{psfrags}
			\psfrag{k}[cc][cc]{$k$}
			\psfrag{m1}[cc][cc]{$m_1$}
			\psfrag{m2}[cc][cc]{$m_2$}
			\psfrag{x1}[cc][cc]{$x_1$}
			\psfrag{x2}[cc][cc]{$x_2$}
			\includegraphics[width=8cm]{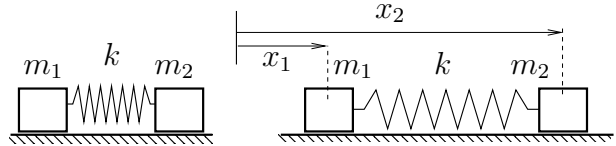}
			\caption{Coupling without resulting external forces.}
			\label{fig1}
		\end{psfrags}
	\end{center}
\end{figure}

Almost all high-level physics textbooks discuss the problem of connecting two point-like bodies of different masses with an ideal spring of constant $k$ (see Fig.~\ref{fig1}). This is probably the simplest mechanical example to illustrate the theorem that, in the absence of external forces, the center of mass (CM) of a mechanical system is stationary (or in uniform motion in a straight line). Interestingly, the generalization of this problem to rotating rigid bodies is much less common (see Fig.~\ref{fig2}).  The author has not encountered this problem in monographs \cite{first}-\cite{third}, problem collections \cite{fourth}-\cite{seventh}, or online platforms. Even if it is assumed that this type of problem has been solved before, the solution is not well-known or easily accessible to physics students. It’s unfortunate because you can learn a lot from these types of exercises. They are neither trivial nor hopelessly difficult, making them ideal for those who want to study mechanics at a higher level. Furthermore, this type of problem can be easily extended and, with a little overkill, can be made as complex as you like. It will be shown that this problem represents a family of coupled systems and is a possible way for students of mechanics to progress from simplicity to complexity.  
Our task forms a kind of node into which several branches of mechanics meet: harmonic oscillation, coupling, rotation of rigid bodies, conservation laws and rolling without slipping. Our aim is to present a detailed solution to this problem. Over the years, the importance of this task has been appreciated by the author and he has come to believe that it can be an effective help to a wide range of people who want to learn mechanics.

The paper is organized as follows. The two point-like bodies problem is discussed briefly in section \ref{absence}, it will be generalized to rotating rigid bodies in section \ref{external}. A system that is not subject to friction forces yet the no-slip condition is still preserved will be introduced in section \ref{stationary}. In section \ref{solutions} in the presence of friction the rotating coupled system is analyzed using various methods. The applicability of the proposed system to teaching is summarized in section \ref{teaching}. Finally, after drawing the conclusion the change of the static friction forces is examined in Appendix A. The calculation of the natural frequencies of the subsystems and their relationship with the Hamiltonian 
is shown in Appendix B and C.

In the following, let us review the two point-like body problem that we mentioned in the introduction and wish to generalize.

\section{Coupling: In the absence of external resultant forces}
\label{absence}

Consider two point-like bodies with masses $m_1$ and $m_2$. The points are connected horizontally with an ideal (massless) spring of elastic constant $k$ and placed on a frictionless horizontal surface. 

The masses are compressed in the direction of the longitudinal axis of the spring and then released. It is easy to prove that the center of mass of the system ($X$) remains at rest. Since the two end points of an ideal spring always exert the same amount of force ($F_1=F_2$), the equation of motion for the two bodies are as follows
\begin{eqnarray}
	k\left(x_2-x_1\right)&=&m_1\ddot{x_1},\label{f1}\\
	-k\left(x_2-x_1\right)&=&m_2\ddot{x_2},\label{f2}
\end{eqnarray}
where $x_1$ and $x_2$ are the coordinates of the bodies. Perhaps the only difficulty for students is deciding which of the spring forces acting on the bodies will be positive/negative. Taking into account the 
$U$ potential energy of the spring 
\begin{equation}
	U=\frac{k}{2}\left(x_2-x_1\right)^2,
\end{equation}
the correct sign can easily be deduced as $F_1=-\partial_{x_1}U$ and $F_2=-\partial_{x_2}U$.
After adding equations (\ref{f1}) and (\ref{f2}) it is given
\begin{eqnarray}
	m_1\ddot{x_1}+m_2\ddot{x_2}&=&0,\\
	(m_1+m_2)\ddot{X}&=&0,
\end{eqnarray}
where $\ddot{X}=\left(m_1\ddot{x_1}+m_2\ddot{x_2}\right)/\left(m_1+m_2\right)$ stands for the acceleration of the CM. After integration over time, the solution corresponding to the initial conditions is if CM of the system remains at rest:
\begin{eqnarray}
	\dot{X}=0.
\end{eqnarray}	
The explanation for this is that the spring forces are internal forces 
and external forces in the vertical direction cancel each other out.

\section{Coupling: involving nonzero resultant external forces}
\label{external}

An obvious generalization of the previous problem is the case when point-like bodies are replaced by rotating rigid bodies. The center of a thin ring of radius $R$ and mass $m_1$ and the center of the cylinder of radius $R$ and mass $m_2$ are connected by an ideal spring with elastic constant $k$. 
It is assumed that the structure of the hoop is similar to that of a bicycle wheel.  Both bodies are placed on a horizontal surface so that the orientation of the spring is also horizontal. It is also assumed that both rigid bodies have a homogeneous mass distribution (the masses of the spokes are negligible) and during the motion they can roll without slipping. No-slip condition is ensured by static friction forces which do not belong to the system. 
\begin{figure}[h!]
	\begin{center}
		\begin{psfrags}
			\psfrag{R}[cc][cc]{$R$}
			\psfrag{k}[cc][cc]{$k$}
			\psfrag{m1}[cc][cc]{$m_1$}
			\psfrag{m2}[cc][cc]{$m_2$}
			\psfrag{x1}[cc][cc]{$x_1$}
			\psfrag{x2}[cc][cc]{$x_2$}
			\psfrag{e1}[cc][cc]{$\varepsilon_1$}
			\psfrag{e2}[cc][cc]{$\varepsilon_2$}
			\psfrag{X}[cc][cc]{$\hat{X}$}
			\psfrag{p1}[cc][cc]{$\varphi_1$}
			\psfrag{p2}[cc][cc]{$\varphi_2$}
			\includegraphics[width=8.5cm]{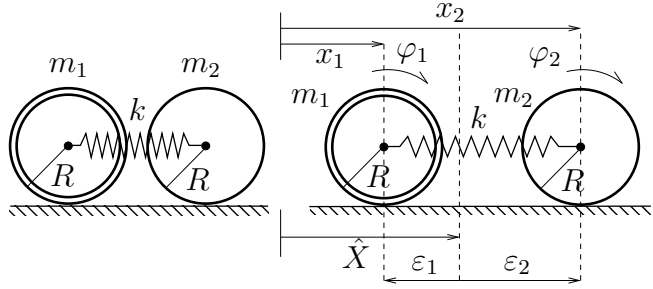}
			\caption{Coupling in the presence of static friction forces.}
			\label{fig2}
		\end{psfrags}
	\end{center}
\end{figure}
As Fig.~\ref{fig2} shows the spring is initially compressed and suddenly released. The system still has two degrees of freedom but because of the constraint relations it is worth using angular coordinates ($\varphi_1$, $\varphi_2$). The relevant constraint relations due to the no-slip condition are as usual:
\begin{subequations}
\begin{equation}
	g_1=x_1-R\varphi_1=0\,\rightarrow\, x_1=R\varphi_1,\\
	g_2=x_2-R\varphi_2=0\,\rightarrow\, x_2=R\varphi_2,
\end{equation}
\end{subequations}

where $x_1$ and $x_2$ represent the coordinates of the centers of the rigid bodies. 
Static friction forces ($f_1$, $f_2$) can be eliminated by the proper choice of the axis of rotation. For this reason they are chosen to be the instantaneous axes (or instant center of rotation) denoted by $z'$. Such an axis passes through the lowest point of the rigid body and parallel with its symmetry axes. The symmetry axis of the cylinder passing through the CM and it is denoted by $z$. The equations for rotational motion for individual bodies are as follows:
\begin{eqnarray}
	F_1R&=&I'_1\ddot{\varphi_1}, \label{e}\\
	-F_2R&=&I'_2\ddot{\varphi_2}, \label{f}	
\end{eqnarray}
where $F_1R$, and $F_2R$ stand for the torques of the spring forces with respect to $z'$ axes.
The spring forces are equal in magnitude and can be expressed as $F=k\left(R\varphi_2-R\varphi_1\right)$.
$I'_1$ and $I'_2$ represent the moment of inertia of the hoop and the cylinder with respect to $z'$ and 
their values due to the Steiner's theorem can be calculated as
\begin{subequations}
\begin{eqnarray}
	I'_1=m_1R^2+m_1d^2=2m_1R^2,\\ 
	I'_2=\frac{1}{2}m_2R^2+m_2d^2=\frac{3}{2}m_2R^2,
\end{eqnarray}
\end{subequations}
where $d=R$ perpendicular distance between the axes $z$ and $z'$. 
By substituting the values of $F$, $I'_1$ and $I'_2$ into equation (\ref{e}) and (\ref{f}) yields
\begin{eqnarray}
	k\left(R\varphi_2-R\varphi_1\right)R&=&2m_1R^2\ddot{\varphi_1}, \label{w2}\\
	-k\left(R\varphi_2-R\varphi_1\right)R&=&\frac{3}{2}m_2R^2\ddot{\varphi_2}, \label{w22}	
\end{eqnarray}
After adding the equations we get
\begin{eqnarray}
	2m_1R^2\ddot{\varphi_1}+\frac{3}{2}m_2R^2\ddot{\varphi_2}=0.
\end{eqnarray}
By integrating over time, the conservation of angular momentum can easily be obtained
\begin{eqnarray}
	I'_1\dot{\varphi_1}+I'_2\dot{\varphi_2}=0 \label{rel1}, \\
	\boldsymbol{L}_1+\boldsymbol{L}_2=0,\,\rightarrow\,\boldsymbol{L}_1=-\boldsymbol{L}_2. \label{g}
\end{eqnarray}
According to the equation (\ref{g}), the rolling bodies oscillate in phase opposition.
This is probably the easiest task to demonstrate the conservation of $\boldsymbol{L}$. Simplifying equation (\ref{rel1}) by $R$ shows that during the oscillation 
$X$ will not be at rest any longer 
\begin{eqnarray}
	2m_1R\dot{\varphi_1}+\frac{3}{2}m_2R\dot{\varphi_2}=0, \label{w}\\
	2m_1\dot{x_1}+\frac{3}{2}m_2\dot{x_2}=0\,\rightarrow\,\dot{X}\neq 0,
\end{eqnarray}
and its role will be taken over by another point ($\hat{X}$).
The location of the new stationary point reads as
\begin{eqnarray}	    
	\hat{X}=\frac{I'_1x_1+I'_2x_2}{I'_1+I'_2}. \label{c}
\end{eqnarray}
The relative coordinates with respect to point $\hat{X}$ are $\varepsilon_1$ and $\varepsilon_2$. We will use them frequently in our later studies. The explanation for the appearance of the new stationary point is that the bodies are subjected to the same spring forces (and their torques), but different amounts of static friction forces ($f_1/f_2=3/2$). The detailed calculation of static friction forces can be found in the \nameref{sec:appendix1}. Intuitively, one might think that this is only because of the different moments of inertia of rigid bodies. And indeed if you connect two thin rings and substitute their moments of inertia into equation (\ref{c})
you get back the position of CM 
\begin{eqnarray}	    
	\hat{X}=\frac{2m_1R^2x_1+2m_2R^2x_2}{2m_1R^2+2m_2R^2}=\frac{m_1x_1+m_2x_2}{m_1+m_2}.
\end{eqnarray}
The same is true for two cylinders. But this is not the whole story. It turns out the situation is more interesting than that. It is shown that, under the right conditions, the steady state of the CM can be achieved for different moments of inertia specifically with a hoop and a cylinder.

\section{How can we make CM stationary?}
\label{stationary}

\begin{figure}[h!]
	\begin{center}
		\begin{psfrags}
			\psfrag{R1}[cc][cc]{$\scriptstyle{R_1}$}
			\psfrag{R2}[cc][cc]{$\small{R_2}$}
			\psfrag{y1}[cc][cc]{$y_1$}
			\psfrag{y2}[cc][cc]{$y_2$}
			\psfrag{k}[cc][cc]{$k$}
			\psfrag{m1}[cc][cc]{$m_1$}
			\psfrag{m2}[cc][cc]{$m_2$}
			\psfrag{p1}[cc][cc]{$\varphi_1$}
			\psfrag{p2}[cc][cc]{$\varphi_2$}
			\includegraphics[width=7cm]{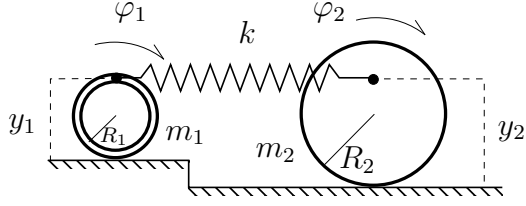}
			\caption{Coupling without friction forces yet ensuring no-slip condition.}
			\label{fig3}
		\end{psfrags}
	\end{center}
\end{figure}

We need to ensure that there is no static friction force on rigid bodies. Caution: the bodies must not slip during rolling. Fortunately, these two conditions can be met in special cases. In general (but not always), the static friction force ensures the no-slip condition during the rolling of rigid bodies. Consider a given cylindrical rigid body with moment of inertia $I$  with respect to its symmetry axis $z$. Assume that the body is subjected to a constant tractive force $F$ in the horizontal direction.
It can be shown that the magnitude and direction of the static friction force depends on the vertical $y$ position (point of application) of $F$. Details can be found in the \nameref{sec:appendix1}.  The change of the static friction force $f(y)$ can be expressed by the following linearly decreasing function
\begin{equation}
	f(y)=F\left(1-\frac{mR}{I+mR^2}\,y\right). \label{fric}
\end{equation}
Knowing this, we are looking for the answer to the question of what $y$ values will give zero static friction force. In the case of a thin ring (hoop), the static friction force turns to zero at the position of the pulling force $y_1=2R$. Similarly, in the case of a homogeneous cylinder the static friction force is eliminated if the position of the pulling force $y_2=3/2R$. For a coupled system, the magnitude of the 
horizontal spring force varies with time, however it seems reasonable to assume that the equation (\ref{fric}) remains valid. Based on the foregoing, we have constructed a coupled system that is not subjected to static friction forces during oscillation yet it does not slip. This system can be seen in Fig.~\ref{fig3}. Since the spring must be horizontal and the centers of rigid bodies are at the same height $R_1=R_2/2$.
Verify that the CM of the coupled system will be stationary during the oscillation. 
It is important to note that the following equations describe the motion of the coupled system only in small angle approximation. This means that the changes in angular coordinates must be small enough for the spring to maintain its horizontal orientation. The relevant dynamical equations with respect to the $z'$ axes of rotation are
\begin{eqnarray}
	F\left(2R_1\right)&=&I'_1\ddot{\varphi_1}, \label{a}\\
	-F\left(\frac{3}{2}R_2\right)&=&I'_2\ddot{\varphi_2} \label{b},	
\end{eqnarray}
where the spring force takes the following form: 
\begin{eqnarray}
	F=k\left(\frac{3}{2}R_2\varphi_2-2R_1\varphi_1\right).
\end{eqnarray}
%\newpage
Dividing equation (\ref{a}) and (\ref{b}) by the corresponding lever arms we obtain
\begin{eqnarray}
	F&=&\frac{I'_1}{2R_1}\ddot{\varphi_1},\\
	-F&=&\frac{I'_2}{\frac{3}{2}R_2}\ddot{\varphi_2},	
\end{eqnarray}
and by adding these equations, the spring forces are eliminated. Thus
\begin{eqnarray}
	\frac{I'_1}{2R_1}\ddot{\varphi_1}+\frac{I'_2}{\frac{3}{2}R_2}\ddot{\varphi_2}=0,\\
	\frac{2m_1R_1^2}{2R_1}\ddot{\varphi_1}+\frac{\frac{3}{2}m_2R_2^2}{\frac{3}{2}R_2}\ddot{\varphi_2}=0,\\
	m_1R_1\ddot{\varphi_1}+m_2R_2\ddot{\varphi_2}=0.
\end{eqnarray}
Finally, taking into account the constraint relations of $x_1=R_1\varphi_1$ and $x_2=R_2\varphi_2$
it is obtained that
\begin{eqnarray}
	m_1\ddot{x_1}+m_2\ddot{x_2}=0,
\end{eqnarray}
and integrating over time, we have 
\begin{eqnarray}
	\label{bb}
	m_1\dot{x_1}+m_2\dot{x_2}=0\,\rightarrow\,\dot{X}=0. 
\end{eqnarray}
As expected, in the absence of friction forces, the CM became stationary again. It is worth noting that although the magnitude of the spring forces acting on the system are equal, but as equations (\ref{a}) and (\ref{b}) shows the magnitudes of their torques are different. Therefore $\boldsymbol{L}$ is not constant of motion in this case.

\section{Other important features of the coupled system}
\label{solutions}

In this section we will derive the equation of motion for the coupled system subjected to non-zero external forces using various methods. Each method shows a slightly different physical characteristic of the problem. For mechanical 
systems that can oscillate and rotate it is worth distinguishing the angular frequency ($\Omega$) 
from the angular speed ($\omega$) in the notation. The equation of motion of an oscillatory system can then  be written as:
\begin{eqnarray}
	\ddot{q}+\Omega^2 q=0,\label{QQ2}
\end{eqnarray}
where $q$ is any relevant coordinate ($x_1$, $x_2$, $\varphi_1$, $\varphi_2$,\dots) and $\Omega=2\pi/T$. 

\subsection{Solution based on the dynamic equations}
\label{s1}

Equations (\ref{w2}-\ref{w22}) describe the motion of the parts of the coupled oscillator, 
but now we are seeking the equation of motion for the entire system.
Therefore, consider again equation (\ref{w}) which, after simplification by $R$ and integration over time, follows
\begin{eqnarray}
	2m_1\varphi_1=-\frac{3}{2}m_2\varphi_2,\,\rightarrow\,\varphi_2=-\frac{4m_1}{3m_2}\varphi_1.
\end{eqnarray}
And finally, substituting $\varphi_2$ into equation (\ref{w2}), we get
the equation of motion for the coupled system
\begin{eqnarray}
	k\left(-R\frac{4m_1}{3m_2}\varphi_1-R\varphi_1\right)R=2m_1R^2\ddot{\varphi_1},\\
	\ddot{\varphi_1}+\frac{k}{2m_1}\left(1+\frac{4m_1}{3m_2}\right)\varphi_1=0.\label{QQ1}
\end{eqnarray}
This equation describes a simple harmonic motion although strictly speaking only the centers of the rolling bodies move in a rectilinear path. 
Comparing equation (\ref{QQ1}) with equation (\ref{QQ2}) gives 
\begin{eqnarray}
	\Omega^2=\left(\frac{k}{2m_1}+\frac{2k}{3m_2}\right) \label{QQQ1}.
\end{eqnarray}
The terms in the angular frequency expression can be identified as the natural frequencies of the subsystems. This will be explained in more detail in subsection \ref{s4} and \nameref{sec:appendix2}.

\subsection{Solution based on conservation of $\boldsymbol{E}$ and $\boldsymbol{L}$}
\label{s2}

The equation of motion for the coupled system can also be obtained by using only its conserved quantities,
specifically with the mechanical energy and angular momentum. It is important to note in our case the rolling occurs in a single plane therefore $\boldsymbol{\omega}$ is parallel to $\boldsymbol{L}$ throughout the motion so $\boldsymbol{L}$ can be considered as a scalar with sign ($\pm L$). Its direction can only be oriented perpendicularly inwards or outwards with respect to the plane of rolling. 
When working with conserved quantities, it is worth choosing the stationary point as the origin of our coordinate system and the relative coordinates can be defined as $\varepsilon_1=x_1-\hat{X}$ and $\varepsilon_2=x_2-\hat{X}$. Their orientation is shown in Fig.~\ref{fig2}. Strictly speaking, it is sufficient to assume the existence of a stationary point, we do not need to know its specific position. Due to the constraint relations on rolling without slipping, these relative coordinates can be expressed in terms of angular coordinates as $\varepsilon_1=R\phi_1$ and  $\varepsilon_2=R\phi_2$. Conservation theorems will be used in our case to calculate the maximum velocities of the centers of the rolling bodies. Knowing these values, the angular frequency can be determined because the kinematics of the harmonic motion tells us that $v_1=A_1\Omega$ and $v_2=A_2\Omega$, where $A_1$ and $A_2$ are the maximum displacements of the subsystems relative to point $\hat{X}$. Since their sum is equal to the maximal deformation of the spring 
$A=A_1+A_2$, therefore $v_1+v_2=A\Omega$ and it implies that $\Omega=(v_1+v_2)/A$. As we have seen, knowing the angular frequency, the equation of motion of the system can be determined. The conservation of $E$ and $L$ are seen as 
\begin{eqnarray}
	\frac{k}{2}A^2=\frac{1}{2}\left(2m_1R^2\right)\dot{\Phi_1}^2+\frac{1}{2}\left(\frac{3}{2}m_2R^2\right)\dot{\Phi_2}^2,
	\label{mu}\\
	0=2m_1R^2\dot{\Phi_1}-\frac{3}{2}m_2R^2\dot{\Phi_2}\label{nu},
\end{eqnarray}
where $A$ represents the amplitude of the resulting oscillation, and $\dot{\Phi_1}, \dot{\Phi_2}$ 
are the maximal angular velocities. It is worth noting that the maximum angular velocities are related to the undeformed state of the spring. Therefore, there is no potential energy term on the right-hand side of equation (\ref{mu}).
Using the instantaneous axes of rotation, the kinetic energies appear to be purely rotational energies.  Similarly, in the case of a given rigid body (e.~g.~a cylinder), the angular momentum associated with its orbital motion ($mR^2\dot{\Phi}$) and spin ($1/2mR^2\dot{\Phi}$) are not separated. The maximum angular velocities can be determined by solving equations (\ref{mu})-(\ref{nu}). For example, it can be expressed $\dot{\Phi_2}=(4m_1/3m_2)\dot{\Phi_1}$ from equation (\ref{nu}) and substitute it into equation (\ref{mu}) then solving the equation for $\dot{\Phi_1}$. Due to the constraint relations associated with the no-slip condition the maximal velocities of the centers of the rolling bodies are $v_1=R\dot{\Phi_1}$ and $v_2=R\dot{\Phi_2}$. Hence
\begin{subequations}
\begin{eqnarray}
	v_1=A\left[\frac{3km_2}{2m_1\left(4m_1+3m_2\right)}\right]^{1/2},\label{om}\\
	v_2=A\left[\frac{8km_1}{3m_2\left(4m_1+3m_2\right)}\right]^{1/2}.\label{pi}
\end{eqnarray}
\end{subequations}
Adding equation (\ref{om}) and equation (\ref{pi}) each other and dividing by $A$ gives
\begin{eqnarray}
	\Omega=\frac{v_1+v_2}{A}=
	\left(\sqrt{\frac{3m_2}{2m_1}}+\sqrt{\frac{8m_1}{3m_2}}\right)\frac{\sqrt{k}}{\sqrt{4m_1+3m_2}}=\nonumber\\
	\left({\frac{3m_2+4m_1}{\sqrt{2m_13m_2}}}\right)\frac{\sqrt{k}}{\sqrt{4m_1+3m_2}}=\nonumber\\
	\left[k\frac{3m_2+4m_1}{2m_13m_2}\right]^{1/2}=\left[\frac{k}{2m_1}+\frac{2k}{3m_2}\right]^{1/2}.
\end{eqnarray}
It is easy to recognize that $\Omega^2$ is identical to equation (\ref{QQQ1}). This method emphasizes that although the maximum velocities of the centers of the bodies include the $A$ amplitude of the motion, the angular frequency (and time period) does not. This phenomenon is unique to harmonic oscillation and is often referred to as isochronism. Since the $v_1$, $v_2$ maximum velocities can be expressed in terms of the $A_1$, $A_2$ amplitudes of the subsystems 
so their values can be determined
\\begin{subequation}
\begin{eqnarray}
	v_1=A_1\Omega\,\rightarrow\,A_1=A\frac{3m_2}{4m_1+3m_2},\\
	v_2=A_2\Omega\,\rightarrow\,A_2=A\frac{4m_1}{4m_1+3m_2}.
\end{eqnarray}
\\end{subequation}
For the same masses $A_1=(3/7)A$ and $A_2=(4/7)A$, so according to the considerations based on conservation of $E$ and $\boldsymbol{L}$ we also came to conclusion that the CM does not remain stationary.

\subsection{Solution based on the Euler-Lagrange equations}
\label{s3}

In the course of the teaching, it is also worth writing down the Lagrangian of the systems and deriving the equations of motion from them. Point $\hat{X}$ has been chosen as the origin of the coordinate system, therefore, the Lagrangian can be written in the following form
\begin{equation}
	\mathcal{L}=\frac{1}{2}I'_1\dot{\phi_1}^2+\frac{1}{2}I'_2\dot{\phi_2}^2-
	\frac{1}{2}k\left(\varepsilon_1+\varepsilon_2\right)^2,
\end{equation}
Interesting to point out if we switch to the new coordinates by working only with dynamic equations 
\begin{eqnarray}
	-k_1\left(R\phi_1\right)R=I'_1\ddot{\phi_1},\label{Q1}\\
	-k_2\left(R\phi_2\right)R=I'_2\ddot{\phi_2},\label{Q2}
\end{eqnarray}
where $k_1$ and $k_2$ are the spring constants of the relevant spring parts, we need to know how the elastic constants of the parts relate to the elastic constant of the whole:
\begin{eqnarray}	
	\frac{1}{k}=\frac{1}{k_1}+\frac{1}{k_2}. \label{part}
\end{eqnarray}
It is intriguing that a solution based on the Euler-Lagrange equations can be derived without this explicit relation.
These equations for the coupled system can be written down as 
\begin{eqnarray}
	\frac{d}{d t}\left(\frac{\partial\mathcal{L}}{\partial\dot{\phi_1}}\right)=\frac{\partial\mathcal{L}}{\partial\phi_1}\rightarrow	I'_1\ddot{\phi_1}=-k\left(R\phi_1+R\phi_2\right)R, \label{al}\\
	\frac{d}{d t}\left(\frac{\partial\mathcal{L}}{\partial\dot{\phi_2}}\right)=\frac{\partial\mathcal{L}}{\partial\phi_2}\rightarrow	I'_2\ddot{\phi_2}=-k\left(R\phi_1+R\phi_2\right)R. \label{be}
\end{eqnarray}
Since the right-hand sides of the equations are equal, it is implied that
\begin{eqnarray}
	2m_1R^2\ddot{\phi_1}=\frac{3}{2}m_2R^2\ddot{\phi_2}, \label{ga}
\end{eqnarray}
after integrating twice over time, we get
\begin{eqnarray}
	2m_1\phi_1=\frac{3}{2}m_2\phi_2\,\rightarrow\,\phi_2=\frac{4m_1}{3m_2}\phi_1 \label{de}.
\end{eqnarray}
And finally, substituting $\phi_2$ into equation (\ref{al}), we get
the differential equation for the oscillation
\begin{eqnarray}
	-k\left(R\phi_1+R\frac{4m_1}{3m_2}\phi_1\right)R=2m_1R^2\ddot{\phi_1},  \label{ep0}\\
	\ddot{\phi_1}+\frac{k}{2m_1}\left(1+\frac{4m_1}{3m_2}\right)\phi_1=0. \label{ep}
\end{eqnarray}
One can see that equation (\ref{ep}) is identical to equation (\ref{QQ1}). 
Finally, calculate the spring constants $k_1$ and $k_2$ of the subsystems. 
To do this, substitute $\phi_1=(3m_2/4m_1)\phi_2$ into equation (\ref{be}), and then we find that
\begin{eqnarray}
	-k\left(R\frac{3m_2}{4m_1}\phi_2+R\phi_2\right)R=\frac{3}{2}m_2R^2\ddot{\phi_2}, \label{pe}
\end{eqnarray}
and comparing equations (\ref{Q1}-\ref{Q2}) with equations 
(\ref{ep0}-\ref{pe}), we conclude that the elastic constants of the subsystems are
\begin{subequations}
\begin{eqnarray}
	k_1=k\frac{3m_2+4m_1}{3m_2}, \label{k1}\\
	k_2=k\frac{3m_2+4m_1}{4m_1}, \label{k2}
\end{eqnarray}
\end{subequations}
thus equation (\ref{part}) is inherently satisfied using the analytical description.
Note $k_1=k_2$ (and $A_1=A_2$) if $m_1/m_2=3/4$, but they are not equal for the same masses:
\begin{eqnarray}
	k_1=\frac{7k}{3},\quad k_2=\frac{7k}{4}.
\end{eqnarray}
We can explain this phenomenon as follows: if $m_1=m_2$ and $x_1=0$ the position of the stationary point is
$\hat{X}=(3/7)l$ which is actually the length of the left spring part. 
The varying length of the spring during the motion is denoted by $l$. Consequently the length 
of the right side spring part is therefore $l-\hat{X}=(4/7)l$. The shorter the spring section, the greater 
its stiffness, therefore $k_1>k_2$. 

\subsection{Solution based on the natural frequencies}
\label{s4}

Our coupled systems performed normal oscillations during their motion.
It means that both parts of the system move sinusoidally with the same frequency and with a fixed phase relation. Consequently the angular frequency squared of the system must be equal to the sum of the natural frequency squared of the two parts. The natural (eigen) frequency means that one of the bodies of the system is imaginatively anchored to the ground but the other is free to oscillate. The resulting vibration is one of the natural frequencies of the coupled system.
To calculate the eigen frequencies in a simple way, the mechanical energy of the subsystem 
can be expressed in quadratic form
\begin{eqnarray}
	E=\frac{1}{2}\alpha\dot{q}^2+\frac{1}{2}\beta q^2.\label{qee}
\end{eqnarray}
If the parameters $\alpha$ and $\beta$ are constants, this quadratic energy expression guarantees the ability of the system to oscillate harmonically, regardless of its construction. As shown in equation (\ref{QQQ1}) the angular frequency square of the coupled system (see Fig.~\ref{fig2}) is 
\begin{eqnarray}
	\Omega_{f}^2=\frac{k}{2m_1}+\frac{2k}{3m_2}\equiv\Omega_1^2+\Omega_2^2,
\end{eqnarray}
where $\Omega_{f}^2$ is indeed the sum of the eigen frequency squares of the individual system components.
The $f$ symbol in the subscript of $\Omega$ indicates that the rigid bodies are subject to friction forces.
The calculation of $\Omega_1$ and $\Omega_2$ natural frequencies based on the quadratic energy expression (equation \ref{qee}) is 
shown in the Appendix B.

\subsection{What the natural frequencies are for?}
As the angular frequencies of the subsystems are easier to calculate than to obtain the equation of motion for the entire system,
knowledge of natural frequencies helps students to detect errors that may occur during the calculation. This applies to both the equation of motion and as we will see the potential energy term of the Hamiltonian.

\subsection{Solution based on the Hamilton's equations}
\label{s5}
This problem also provides an excellent opportunity to apply canonical variables ($p$, $q$).
Worth to compare the Lagrangian of the system with its Hamiltonian 
\begin{eqnarray}
	\mathcal{L}=m_1R^2\dot{\phi_1}^2+\frac{3}{4}m_2R^2\dot{\phi_2}^2-
	\frac{k}{2}\left(R\phi_1+R\phi_2\right)^2,\label{l}\\
	\mathcal{H}=\frac{p_1^2}{2m_1}+\frac{p_2^2}{2m_2}+\frac{k}{2}\left(\frac{q_1}{\sqrt{2}}+\frac{\sqrt{2}q_2}{\sqrt{3}}\right)^2,
	\label{k}
\end{eqnarray}
where ($p_1$,$q_1$) and ($p_2$,$q_2$) are the canonical coordinates.  
The relationship between the ($p$,$q$) and ($\dot{\phi}$,$\phi$) coordinates is obtained by equating the first two terms of equation (\ref{l}) and equation (\ref{k}) with each other. As a result of this we get $q_1=\sqrt{2}R\phi_1$ and $q_2=\sqrt{3/2}R\phi_2$. 
Strictly speaking, the potential energy term of Hamiltonian can only be written down after these relations have been recognized. 
In the Appendix C we will show that the coefficients of the $q_1$, $q_2$ coordinates of the potential energy term can be also determined by knowing the natural frequencies of the subsystems. For the coupled oscillator, the relevant Hamilton's equations can be written as:
\begin{eqnarray}
	-\dot{p_1}=\frac{\partial\mathcal{H}}{\partial q_1}\,\rightarrow\,	-\dot{p_1}=k\left(\frac{q_1}{\sqrt{2}}+
	\frac{\sqrt{2}q_2}{\sqrt{3}}\right)\frac{1}{\sqrt{2}},\label{h}\\
	-\dot{p_2}=\frac{\partial\mathcal{H}}{\partial q_2}\,\rightarrow\,	-\dot{p_2}=k\left(\frac{q_1}{\sqrt{2}}+
	\frac{\sqrt{2}q_2}{\sqrt{3}}\right)\frac{\sqrt{2}}{\sqrt{3}}.
\end{eqnarray}
It is easy to see that the right-hand sides of the two equations become equal if equation (\ref{h}) is multiplied by $2/\sqrt{3}$ then making the left-hand sides of the equations equal:
\begin{eqnarray}
	\dot{p_2}=\dot{p_1}\frac{2}{\sqrt{3}}\,\rightarrow\,q_2=\frac{m_1}{m_2}\frac{2}{\sqrt{3}}q_1.
\end{eqnarray}
Substituting $q_2$ into equation (\ref{h}) gives the equation of motion for the coupled system in canonical coordinates
\begin{eqnarray}
	-\dot{p_1}=k\left(\frac{q_1}{2}+
	\frac{2m_1}{3m_2}q_1\right).
\end{eqnarray}
Since $\dot{p_1}=m_1\ddot{q_1}$ this equation of motion is equivalent to equation (\ref{ep}).
Worth noting that in our case the other two Hamilton's equations
\begin{subequations}
\begin{eqnarray}
	\dot{q_1}=\frac{\partial\mathcal{H}}{\partial p_1}\,\rightarrow\,\dot{q_1}=\frac{p_1}{m_1},\\
	\dot{q_2}=\frac{\partial\mathcal{H}}{\partial p_2}\,\rightarrow\,\dot{q_2}=\frac{p_2}{m_2},	
\end{eqnarray}
\end{subequations}
do not provide any new information about the dynamics of the system. 
It is also worth mentioning that in the case of considering the vibrations of two point-like bodies (see Fig.~\ref{fig1}), there is little difference in the Lagrangian and Hamiltonian of the system because the coordinates in $U$ term of the Lagrangian and the Hamiltonian are interchangeable. For this reason, solving our central task allows students to learn more effectively about the application of Hamilton's equations.
\begin{figure*}[hbt!]
	\begin{center}
		\begin{psfrags}
			\psfrag{a}[cc][cc]{$\,\,(a)$}
			\psfrag{b}[cc][cc]{$\,(b)$}
			\psfrag{c}[cc][cc]{$\,(c)$}
			\psfrag{k}[cc][cc]{$k$}
			\psfrag{R}[cc][cc]{$R$}
			\psfrag{3R}[cc][cc]{$3R$}
			\psfrag{R2}[cc][cc]{$3R$}
			\psfrag{m}[cc][cc]{$m$}
			\psfrag{x1}[cc][cc]{$x_1$}
			\psfrag{x2}[cc][cc]{$x_2$}
			\psfrag{P1}[cc][cc]{$\varphi_1$}
			\psfrag{P2}[cc][cc]{$\varphi_2$}
			\psfrag{k}[cc][cc]{$k$}
			\psfrag{m1}[cc][cc]{$m_1$}
			\psfrag{m2}[cc][cc]{$m_2$}
			\psfrag{p1}[cc][cc]{$\varphi_1$}
			\psfrag{p2}[cc][cc]{$\varphi_2$}
			\includegraphics[width=15cm]{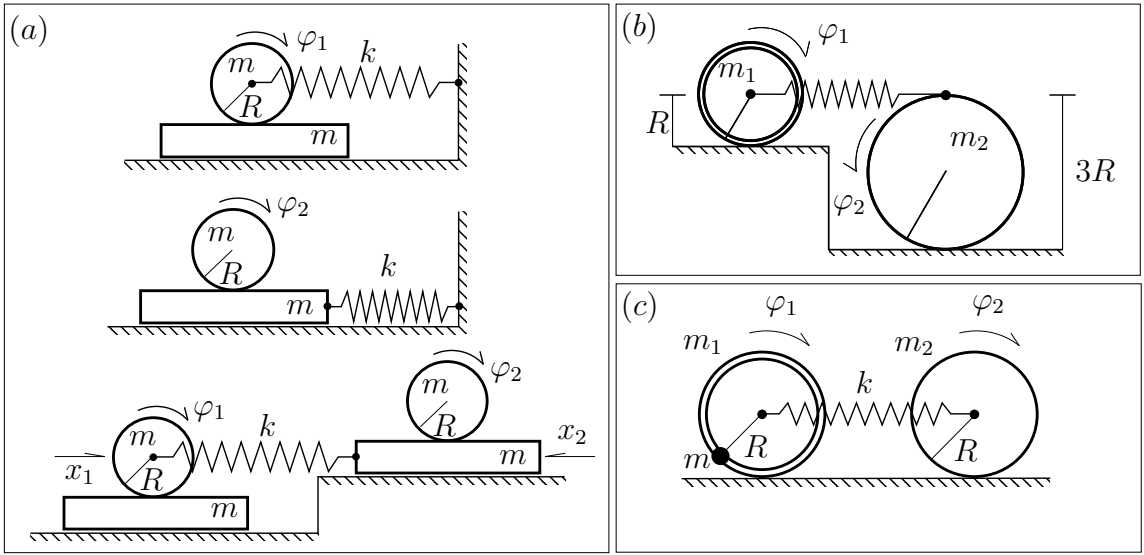}
			\caption{Part $(a)$ and $(b)$: A gradual, step-by-step build-up of complexity. 
				Part $(c)$: Increasing difficulty in leaps and bound. }
			\label{fig4}
		\end{psfrags}
	\end{center}
\end{figure*}
\FloatBarrier

\section{How should our approach be used in teaching?}
\label{teaching}

Our proposed problem can be an effective help for those students who are in the learning phase of advanced mechanics.
We offer a simple problem that is complex enough to demonstrate the following principles: simple harmonic motion, coupling, conservation of mechanical energy, conservation of angular momentum, rolling of rigid bodies. We offer an easily digestible problem to show how Hamilton's equations work. Based on the author's reading experience the usual tasks discussed in textbooks are too abstract for a beginner to cope with. Our system bridges the gap between simple and complex problems. Our problem is an excellent training ground for different solving techniques. It is a huge advantage for problem solvers to know in advance what you have to come up with (besides, such solutions are also entertaining). The complexity of the problem is flexible and its configuration can be easily changed. For example, by changing the vertical centers of the rigid bodies and the anchorage points of the springs. As it can be seen in Fig.~\ref{fig4} $(a)$ and $(b)$, complexity in this type of problem can be built up gradually, step-by-step. In the author's experience, the improved exercises can be used effectively as project work.

\section{Conclusion}
It was shown that during normal modes of an oscillatory system consisting of a hoop and a cylinder joining their centers by an ideal spring, its CM does not remain at rest. This effect is due to the resultant external forces acting on the system which disappears when the coupled rigid bodies have the same moment of inertia. However, in case of different moment of inertia by proper positioning the two fixed end points of the spring vertically, it was shown that the CM of the system remains at rest. The equation of motion of the coupled system has also been derived using dynamic equations, Lagrange's equations, Hamilton's equations and even by applying the conservation laws of energy and angular momentum.  It has been pointed out that the natural frequencies of the subsystems are easy to calculate and therefore provide a way to check the equations of motion obtained by other methods. Several didactical advantages in teaching have been discussed. The simplicity of the problem makes the rich physical background unexpected. The reason for this is it lies on a narrow borderline between simplicity and complexity.

\appendix
\subsection*{Appendix A: Variation of static friction forces}
\label{sec:appendix1}

Consider our central problem, shown in Fig.~\ref{fig2}. We will first show how the static friction forces vary with the displacement of the centers of the rolling bodies. Point $\hat{X}$ has been selected as the origin of our coordinate system and the relative coordinates are 
defined as $\varepsilon_1=x_1-\hat{X}$ and $\varepsilon_2=x_2-\hat{X}$. Their orientation is shown in Fig.~\ref{fig2} and
we can introduce angular coordinates by taking into account the constraint relations as $\varepsilon_1=R\phi_1$ and $\varepsilon_2=R\phi_2$.
Since the angular acceleration of a rigid body cannot depend on the choice of different axes of rotation ($z'$ and $z$), so the torques of the spring forces and friction forces must generate the same angular accelerations. Using this fact, friction forces can be expressed and compared with each other. The equations of motions for the rigid bodies with respect to the $z'$ and $z$ axes are:
\appendix
\setcounter{section}{1}
\begin{eqnarray}
	-k_1\left(R\phi_1\right)R=I'_1\ddot{\phi_1}\,&\rightarrow\,\ddot{\phi_1}&=-\frac{k_1}{2m_1}\phi_1,\\
	-f_1R=I_1\ddot{\phi_1}\,&\rightarrow\,\ddot{\phi_1}&=-\frac{f_1}{m_1R},\\	
	-k_2\left(R\phi_2\right)R=I'_2\ddot{\phi_2}\,&\rightarrow\,\ddot{\phi_2}&=-\frac{2k_2}{3m_2}\phi_2,\\
    -f_2R=I_2\ddot{\phi_2}\,&\rightarrow\,\ddot{\phi_2}&=-\frac{2f_2}{m_2R}.
\end{eqnarray}
Eliminating the angular accelerations in pairs and substituting the values of $k_1$ and $k_2$ from equations (\ref{k1}) and (\ref{k2}), 
we obtain that
\begin{eqnarray}
	f_1=\frac{k_1}{2}R\phi_1=k\frac{3m_2+4m_1}{6m_2}R\phi_1,\\
	f_2=\frac{k_2}{2}R\phi_2=k\frac{3m_2+4m_1}{12m_1}R\phi_2.
\end{eqnarray}
Using equation (\ref{de}) which gives $\phi_1/\phi_2=3m_2/4m_1$ the ratio of the static friction forces is
\begin{eqnarray}
	\frac{f_1}{f_2}=\frac{3}{2}\,\rightarrow\,f_1>f_2
\end{eqnarray}	
as expected.
\begin{figure}[hbt!]
	\begin{center}
		\begin{psfrags}
			\psfrag{P1}[cc][cc]{$\ddot{\varphi}$}
			\psfrag{R}[cc][cc]{$R$}
			\psfrag{A}[cc][cc]{$A$}
			\psfrag{B}[cc][cc]{$B$}
			\psfrag{I}[cc][cc]{$I$}
			\psfrag{I2}[cc][cc]{$f_1(y)$}
			\psfrag{I3}[cc][cc]{$f_2(y)$}
			\psfrag{o}[cc][cc]{$0$}
			\psfrag{m}[cc][cc]{$m$}
			\psfrag{y}[cc][cc]{$y$}
			\psfrag{R2}[cc][cc]{$\scriptstyle{2R}$}
			\psfrag{R3}[cc][cc]{$\frac{3R}{2}$}
			\psfrag{P}[cc][cc]{$\varphi_1$}
			\psfrag{F}[cc][cc]{$\boldsymbol{F}$}
			\psfrag{F2}[cc][cc]{$F$}
			\psfrag{f}[cc][cc]{$\boldsymbol{f}$}
			\psfrag{F3}[cc][cc]{$-\frac{F}{3}$}
			\psfrag{f2}[cc][cc]{$f(y)$}
			\includegraphics[width=7cm]{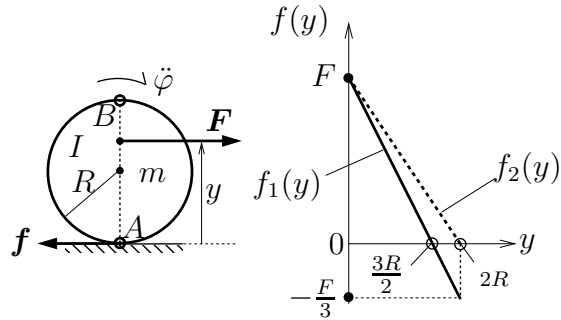}
			\caption{Left: the relevant mechanical system. Right: variation of static friction force as a function of $y$.}
			\label{fig5}
		\end{psfrags}
	\end{center}
\end{figure}

In the following we show how the magnitude and direction of the static friction force depends on the vertical $y$ position (point of application) of $F$. As Fig.~\ref{fig5} shows a cylindrical rigid body of mass $m$, radius $R$, and moment of inertia $I$ with respect to its symmetry axis is placed on a horizontal surface. A tractive force $\boldsymbol{F}$ of constant magnitude is applied in the horizontal direction, such that its point of application can only vary vertically along the $AB$ axis. It is assumed that the rigid body cannot slip during motion and the variation of $\varphi$ is infinitesimally small. 
The dynamic equations for the rigid body in small angle approximation are 
\begin{eqnarray}
	F-f=ma, \label{a1}\\
	Fy=\left(I+mR^2\right)\ddot{\varphi}\label{a2},
\end{eqnarray}
where $f$ stands for the static friction force and $a$ is the translational acceleration of CM. Exploiting the constraint relation $a=R\ddot{\varphi}$ due to the no-slip condition the angular acceleration can be expressed from both (\ref{a1}) and (\ref{a2}) equations, so
\begin{eqnarray}
	\ddot{\varphi}=\frac{F}{mR}-\frac{f}{mR},\\
	\ddot{\varphi}=\frac{F y}{\left(I+mR^2\right)}.
\end{eqnarray}
By eliminating $\ddot{\varphi}$ it is given 
\begin{eqnarray}
	\frac{F}{mR}-\frac{f}{mR}=\frac{F y}{\left(I+mR^2\right)},
\end{eqnarray}
and finally we get
\begin{eqnarray}	
	f\left(y\right)=F-\frac{FmR}{\left(I+mR^2\right)}y.
\end{eqnarray}
This linearly decreasing function is plotted in Fig.~\ref{fig5} for both cylinder ($f_1(y)$) and hoop ($f_2(y)$).
It is worth noting that in the case of a cylinder, if $y>1.5R$ the orientation of $\boldsymbol{f}$ is reversed (pointing to the right), whereas in the case of a hoop its orientation is unchanged (always points left). Furthermore, in the case where $y=0$, the magnitude of the static friction force (irrespective of the current value of $I$) is equal to $F$, so the acceleration of the rigid body is zero. Finally, it should be stressed that the static friction force not being able to do work if its point of application is stationary.

\subsection*{Appendix B: Calculation of the natural frequencies}
\label{sec:appendix2}
\appendix
\setcounter{section}{2}
Consider our central problem, shown in Fig.~(\ref{fig2}). Its first subsystem consists of a hoop of radius $R$ and mass $m_1$ and a horizontal spring with elastic constant $k$. Its mechanical energy can be written down as $E_1=\left(I'_1/2\right)\dot{\phi_1}^2+k/2(R\phi_1)^2$, so the relevant parameters and $\Omega_1$ are as follows:
\begin{eqnarray}	
	\alpha_1=2m_1R^2,\quad \beta_1=kR^2,\\
	\Omega_1=\sqrt{\frac{\beta_1}{\alpha_1}}=\sqrt{\frac{kR^2}{2m_1R^2}}=\sqrt{\frac{k}{2m_1}}.
\end{eqnarray}	
Our second subsystem consists of a cylinder of radius $R$ and mass $m$ and a horizontal spring with elastic constant $k$. 
Its mechanical energy can be written down as $E_2=\left(I'_2/2\right)\dot{\phi_2}^2+k/2(R\phi_2)^2$, so the relevant parameters and $\Omega_2$ are seen:
\begin{eqnarray}	
	\alpha_2=\frac{3}{2}m_2R^2,\quad \beta_2=kR^2,\\
	\Omega_2=\sqrt{\frac{\beta_2}{\alpha_2}}=\sqrt{\frac{kR^2}{3/2m_2R^2}}=\sqrt{\frac{2k}{3m_2}}.
\end{eqnarray}	
As shown in (\ref{QQQ1}) the angular frequency square of the coupled system (see Fig.~\ref{fig2}) is 
\begin{eqnarray}
	\Omega_{f}^2=\frac{k}{2m_1}+\frac{2k}{3m_2},
\end{eqnarray}
where $\Omega_{f}^2$ is indeed the sum of the eigen frequency squares of the individual system components.
The $f$ symbol in the subscript of $\Omega$ indicates that the rigid bodies are subject to friction forces.
Based on completely similar considerations, the angular frequency square of the rolling system (shown in Fig.~\ref{fig3}) without static friction is
\begin{eqnarray}
	\Omega_{\bcancel{f}}^2=\frac{2k}{m_1}+\frac{3k}{2m_2}\equiv\Omega_3^2+\Omega_4^2.
\end{eqnarray}

\subsection*{Appendix C: Relationship with the Hamiltonian}
\label{sec:appendix3}

We show how the potential energy term of the Hamiltonian related to the natural frequencies of the subsystems.
Considering the Hamiltonian of the coupled system 
\appendix
\setcounter{section}{3}
\begin{eqnarray}
	\mathcal{H}_{f}=\frac{p_1^2}{2m_1}+\frac{p_2^2}{2m_2}+\frac{k}{2}\left(\frac{q_1}{\sqrt{2}}+\frac{\sqrt{2}q_2}{\sqrt{3}}\right)^2.
	\label{k0}
\end{eqnarray}
 In the potential energy term of the Hamiltonian, the coefficients $1/\sqrt{2}$ and $\sqrt{2/3}$ of coordinates $q_1$, $q_2$ can also be obtained from the normal frequencies of the subsystems. More specifically
\begin{eqnarray}
	U_{f}=\frac{k}{2}\left(\frac{\Omega_1}{\sqrt{k/m_1}}\,q_1+\frac{\Omega_2}{\sqrt{k/m_2}}\,q_2\right)^2,
	\label{k0}
\end{eqnarray}
where $\Omega_1=\sqrt{k/2m_1}$ and $\Omega_2=\sqrt{2k/3m_2}$. On the basis of considerations entirely analogous to the above for a coupled system without friction forces (shown in Fig.~\ref{fig3}) its Hamiltonian potential energy term seems  
\begin{eqnarray}
	U_{\bcancel{f}}=\frac{k}{2}\left(\frac{\Omega_3}{\sqrt{k/m_1}}\,q_3+\frac{\Omega_4}{\sqrt{k/m_2}}\,q_4\right)^2, \label{k02}
\end{eqnarray}
where $\Omega_{3}=\sqrt{2k/m_1}$ and $\Omega_4=\sqrt{3k/2m_2}$.

\end{document}